\documentstyle[11pt,amsfonts]{article}

\newcommand{\be}{\begin{equation}}
\newcommand{\ee}{\end{equation}}
\newcommand{\bea}{\begin{array}}
\newcommand{\ea}{\end{array}}
\newcommand{\beqa}{\begin{eqnarray}}
\newcommand{\eeqa}{\end{eqnarray}}
\newcommand{\bean}{\begin{eqnarray*}}
\newcommand{\eean}{\end{eqnarray*}}

\def\up#1{\leavevmode \raise.16ex\hbox{#1}}

\def\sqr#1#2{{\vcenter{\vbox{\hrule height.#2pt
        \hbox{\vrule width.#2pt height#1pt \kern#1pt
          \vrule width.#2pt}
        \hrule height.#2pt}}}}

\newcommand{\gapproxeq}{\lower .7ex\hbox{$\;\stackrel{\textstyle
>}{\sim}\;$}}
\newcommand{\lapproxeq}{\lower .7ex\hbox{$\;\stackrel{\textstyle
<}{\sim}\;$}}

\newcounter{appendice}

\def\thebibliography#1{{\bf REFERENCES\markboth
 {REFERENCES}{REFERENCES}}\list
 {[\arabic{enumi}]}{\settowidth\labelwidth{[#1]}\leftmargin\labelwidth
 \advance\leftmargin\labelsep
 \usecounter{enumi}}
 \def\newblock{\hskip .11em plus .33em minus -.07em}
 \sloppy
 \sfcode`\.=1000\relax}

\begin{document}

%
%

\vspace*{5mm}
\centerline{ \LARGE Edge States from}
\bigskip
\centerline{ \LARGE  Defects on the Noncommutative
  Plane\footnote{Talk presented by second author at the conference ``Space-time and
  Fundamental Interactions:  Quantum Aspects'' in honor of
  A.P. Balachandran's $65$th birthday, Vietri sul Mare, Salerno, Italy
  $26$th-$31$st May, 2003.}}

\vskip 2cm

\centerline{ {\sc A. Pinzul\footnote{Address after September 2003:
Physics Department, Syracuse University, Syracuse, NY 13244-1130,
USA.} and A. Stern}  }

\vskip 1cm
\begin{center}
Department of Physics, University of Alabama,\\
Tuscaloosa, Alabama 35487, USA
\end{center}

\begin{abstract}

We illustrate how  boundary states are  recovered when going from a
noncommutative manifold to a commutative one with a boundary.  Our example is
the noncommutative plane with a defect, whose 
commutative limit was found to be a
punctured plane - so here the boundary is one point.    Defects were introduced by removing states
from the standard harmonic oscillator Hilbert space.  For  Chern-Simons
theory, the defect acts as a source, which was found to be  associated with a nonlinear
deformation of the $w_\infty$ algebra.  The undeformed $w_\infty$
algebra is recovered in
the commutative limit, and here we show that its spatial support is in a tiny region near the puncture.

\end{abstract}


\newpage

\section{Introduction}
\setcounter{equation}{0}

Field theory on the noncommutative plane has been well studied.\cite{reviews}   It
is a
nontrivial deformation of field theory on the commutative plane and
has implications for strings, gravity, renormalization and the
fractional quantum Hall effect (FQHE).
There are good reasons for finding analogous deformations of other
commutative manifolds.   Of particular interest, especially for
topological theories, 
are  manifolds with boundaries.  A primary  example is  Chern-Simons theory,
which  is  empty on the
plane.  The same applies for   Chern-Simons theory on the
noncommutative plane.   The theory becomes nontrivial when written on
a commutative manifold with a boundary,  and all the  dynamics
resides  at the boundary.\cite{wit},\cite{bbgs}   These are so-called
edge states, a familiar phenomena in the FQHE.   It is then natural to search for noncommutative manifolds
which in some limit reduce to commutative manifolds with boundaries, and to demonstrate that
edge states are recovered in the limit.  This is of physical interest
because of the proposal by Susskind that noncommutative $U(1)$  Chern-Simons
theory may describe short distance effects in the FQHE.\cite{sus}  
Another exciting possibility is the application to $2+1$ quantum gravity which can be
written in terms of Chern-Simons theory based on the Poincar\'e gauge group.\cite{wit2}  

The  manifold we consider  is one with a very simple boundary.  It is the plane with one
point removed, where the puncture possibly represents a quasiparticle in the FQHE or a
conical singularity in $2+1$ gravity.  In ref. 
\cite{nnsps}  we showed that the
noncommutative analogue of this manifold is obtained after  removing states
from the infinite dimensional harmonic oscillator Hilbert space of the
noncommutative plane.  In ref. \cite{Pinzul:2002fi}  we wrote down Chern-Simons
theory on this manifold, and showed that the gauge invariant
observables form a nonlinear
deformation of the $w_\infty$ algebra.   The `undeformed' $w_\infty$ algebra is
associated with area preserving diffeomorphisms, and its  relevance  for 
FQHE phenomena is known.\cite{IKS}  So if the proposal of ref. \cite{sus} is
correct the  nonlinear
deformation may provide a more accurate description for the FQHE.
The `undeformed'   $w_\infty$ algebra is
recovered in the commutative limit.   Here we show
that its spatial support is in a shrinking
region near the puncture.  Therefore  we recover edge
states in the limit.

 There has been an effort to write  Chern-Simons systems on the space
 of  finite matrices  for the purpose of describing a FQHE
 droplet.\cite{poly}  It
 was found that for these systems the
question of taking the commutative limit and recovering systems with
boundaries and edge states is subtle.\cite{us2}    Recent progress has
 been made in \cite{Balachandran:2003vm} and  \cite{Lizzi:2003ru}.
 The approach taken here  may also be adapted in this direction.\cite{Pinzul:2002fi}

In section 2 we review the noncommutative plane, while defects are
inserted in section 3.  Chern-Simons theory is written on the
noncommutative plane in sec. 4 and on the noncommutative plane with
defect in sec. 5 where we recover edge states.  
Although this report relies heavily on calculations in refs.
\cite{nnsps},  \cite{Pinzul:2002fi},  \cite{gps} we make it as
self-contained as space allows.

\section{ The  noncommutative plane }

  The   noncommutative plane ${\cal M}^{(0)}$ is generated by the Heisenberg algebra.  Say operator
$z$ and its  hermitian conjugate  $ z^\dagger $ satisfy the commutation
relation 
 \begin{equation}
 [z, z^\dagger] = \Theta_0 \label{zbz}\;,
\end{equation}
 where $\Theta_0$ is a dimensionful nonzero central element, and is identified with
 the noncommutativity parameter.   By rescaling $z$ and
$ z^\dagger$ one gets standard annihilation and
creation operators ${\bf a}$ and ${\bf a}^\dagger$, $z = \sqrt{\Theta_0}\; {\bf a} $ and
$z^\dagger = \sqrt{\Theta_0} \;{\bf a}^\dagger $, which can be realized  on the harmonic
oscillator Hilbert space   $ H^{(0)}$, having basis
$\{|n>,\;n=0,1,2,...\}$, with $\;{\bf a}|0>=0$ and hence  $z|0>=0$.  A field
$\Phi$ on the noncommutative plane  is a polynomial function of $z$  and
$ z^\dagger$.  A pair of commuting
inner derivatives $\nabla_z$ and  $\nabla_{z^\dagger}$ can be defined
\begin{equation}
\nabla_z = -i[p,\; ]\;,\qquad  \nabla_{z^\dagger} = -i[p^\dagger,\; ]
\label{cod}
\end{equation}
The requirement that they commute means that the commutator of the
operator $p$ with its hermitian conjugate $p^\dagger$ is a central
element.  We once again get a Heisenberg algebra, which
expressable in terms of ${\bf a}$ and ${\bf a}^\dagger$.  A convenient choice is \begin{equation}
p =-i\Theta_0^{-1/2}\; {\bf a} ^\dagger\;,\qquad p^\dagger =i\Theta_0^{-1/2}\;
{\bf a} \;,\label{dfp}\end{equation}
 for then  $\nabla_z$
and  $\nabla_{z^\dagger}$ resemble ordinary partial derivatives:
$\nabla_zz=\nabla_{z^\dagger}z^\dagger = 1$, $\nabla_zz^\dagger=\nabla_{z^\dagger}z = 0$.

Using standard coherent states the  product between functions on the
noncommutative plane can be mapped to
an associative star product between functions on the commutative plane.
The  standard coherent states $\{|\zeta>\;, \zeta\in{\mathbb{C}}\} $ form an overcomplete basis of unit
vectors with a resolution of unity which diagonalize ${\bf a}$, and
hence $z$
\begin{equation} z |\zeta> =\zeta
|\zeta>\label{diagz} \end{equation}  They can be obtained by
acting on the ground state $|0>$ with the unitary operator
$U(\zeta,\bar \zeta) = \exp{\;i(\zeta p +\bar \zeta p^\dagger )}$.   A field
$\Phi$ on the noncommutative plane  is mapped to a function $\phi$ on
the complex plane (called the `covariant symbol' of $\Phi$) according to
\begin{equation}\Phi \rightarrow \phi(\zeta,\bar
  \zeta)=<\zeta|\Phi|\zeta>\;,
 \end{equation} while
the operator product between two fields $\Phi$ and $\Psi$ is mapped to
the star product
\begin{equation}
\Phi\Psi \rightarrow [\phi\star\psi](\zeta,\bar \zeta)=<\zeta|\Phi\Psi|\zeta>
\label{dfsp}\end{equation}
It can be expressed compactly in terms of an infinite number of derivatives
\begin{equation}\star\;\;=\exp{\;\Theta_0 \;\overleftarrow{ \frac\partial{ \partial\zeta }}\;\;
\overrightarrow{ {\partial\over{ \partial\bar\zeta} }}}\;,\label{eefsp}\end{equation}
and  is called the  Voros star product\cite{Voros}, which is equivalent
to the Moyal  star
product.\cite{Moyal}  The lowest order terms in (\ref{eefsp}) define
the commutative limit $\Theta_0\rightarrow 0$.  In this limit, the star
product between functions reduces to the point-wise product, while the
star commutator goes to the Poisson bracket:
 \begin{eqnarray}
 [\phi\star\psi](\zeta,\bar \zeta)&\rightarrow &  \phi(\zeta,\bar \zeta)\;
\psi(\zeta,\bar \zeta) \cr &&\cr
[\phi\star\psi-\psi\star\phi](\zeta,\bar \zeta)&\rightarrow &
i\Theta_0\;  \{ \phi,\psi\}(\zeta,\bar \zeta)\;,\label{pwp} \end{eqnarray} where
\begin{equation}
  \{ \phi,\psi\}(\zeta,\bar \zeta) =
-i  \;\biggl( \frac{\partial\phi}{ \partial\zeta }\;\;
 \frac{\partial \psi}{ \partial\bar\zeta} \;-\;
 \frac{\partial\phi}{ \partial\bar\zeta }\;
\frac {\partial\psi}{ \partial\zeta} \biggr)\end{equation}

\section{ The  noncommutative plane with defect}

In ref. \cite{nnsps} defects were inserted on the    noncommutative plane by
removing states from   $ H^{(0)}$.  For example, consider projecting out
the first $n_0$ states $|0>,|1>,...,|n_0 -1>$, and call the result
$ H^{(n_0)}$.   This should be the Hilbert space on which generators of 
the new noncommuting manifold   ${\cal M}^{(n_0)}$  act.  For convenience we again denote
 generators by $z$ and   $ z^\dagger $, and assume they behave
as  lowering and raising operators, respectively.  So we now  need that $z|n_0>=0$.  If we
consider $ H^{(n_0)}$ embedded in $ H^{(0)}$, then the 
lowering and raising operators ${\bf a}$ and ${\bf a}^\dagger$ acting on the latter
cannot be expressed in terms of the lowering and  raising  operators  $z$
and   $ z^\dagger $ acting on the
former.  Moreover,
(\ref{zbz}) cannot be realized on $ H^{(n_0)}$.  We can still define a pair of commuting inner derivatives as in (\ref{cod}),
with $p$ and $ p^\dagger$ written again in terms of ${\bf a}$,
${\bf a}^\dagger$ and the   noncommutativity parameter  $\Theta_0$,
as in (\ref{dfp}), even though $p$ is not well defined
on   $ H^{(n_0)}$.  In order that
$\nabla_z$ and  $\nabla_{z^\dagger}$ are well defined on ${\cal
  M}^{(n_0)}$ we have to impose appropriate boundary conditions 
on the fields.    A field
$\Phi$ is once again a polynomial function of $z$  and
$ z^\dagger$.  It has a well defined  action on  $ H^{(n_0)}$, or
alternatively we can set $\Phi|n_0-1>=\Phi|n_0-2> =...=\Phi|0> =
0$.   If $\nabla_z\Phi$ is to have a well defined action on   $
H^{(n_0)}$ we also need $\Phi|n_0> =0$.  (Similarly, if $\nabla_{z^\dagger}\Phi$ is
to have a well defined action on the dual of  $
H^{(n_0)}$ we need $<n_0|\Phi =0\;$.)      Stronger boundary
 conditions are needed for higher derivatives  to be defined.

As stated above, (\ref{zbz}) cannot be realized on $ H^{(n_0)}$.
Rather one can write 
 \begin{equation}
 [z, z^\dagger] = \Theta (z z^\dagger)
\end{equation}
The  function $
\Theta$ is assumed to be nonsingular on $H^{(n_0)}$, and can be determined after re-expressing $p$ and $p^\dagger$ in
terms of generators  $z$ and   $ z^\dagger $.  A natural choice is
\begin{equation}
p =-i z^\dagger\;\Theta(z z^\dagger)^{-1}\;,\qquad p^\dagger
=i\Theta(z z^\dagger)^{-1} \;z \;,\label{pitz}\end{equation} as it reduces
correctly to the case of the noncommutative plane when
$\Theta=\Theta_0 $.  More generally, after identifying\footnote{There is a subtlety due to the fact that
  the action of $p^\dagger$   on  the ground state
$|n_0>$   differs for   (\ref{dfp})
and (\ref{pitz}), as ${\bf a}$ takes $|n_0>$ out of $
H^{(n_0)}$.  However, due to the boundary conditions on the fields, the
derivatives obtained with either  (\ref{dfp})
or (\ref{pitz}) can be identified.} (\ref{dfp})
and (\ref{pitz}) a recursion relation  was found for $\Theta$ in
ref. \cite{nnsps}.  From dimensional arguments $\Theta$ is linear in  
 $\Theta_0$.  Also it has the feature
that for states with increasing eigenvalue of the  number operator one
approaches a constant value of $\Theta$, and thus the noncommutative
plane.  The behavior is
\begin{equation}   
 \Theta (z z^\dagger)^{-1}|n> \rightarrow \Theta_0^{-1}\;\biggl(1 +
 \frac {r_0}{\sqrt{\Theta_0\;n}}\biggr)|n> \;,\qquad {\rm as}\qquad n\rightarrow\infty\;,\label{lnl}\end{equation}
where $r_0$ is a dimensionful constant depending on $n_0$.

  The  large $n$ limit corresponds to the
 limit of large distances.  This statement can be made more precise
 after again introducing coherent states.  In this regard, the
 standard coherent states used in the previous section are not
 convenient for this purpose because they no longer diagonalize $z$.  An alternative
 set of coherent states (which for convenience we again denote by
 $\{|\zeta>\}$) were developed in ref.  \cite{mmsz} which preserve
 (\ref{diagz}).  Like the  standard coherent states, they form an overcomplete basis of unit
vectors with a resolution of unity.  On the other hand, they are not obtained by
acting on the ground state (now $|n_0>$) with a unitary operator.
 In ref. \cite{gps} we used this set of coherent states to construct a
 star product, and showed that it reduces to the Voros star product in
 the limit that $\Theta$ goes to a central element.  The construction
 is the same as in (\ref{dfsp}), although now this does not result
 in a compact expression for  $\star$ like in (\ref{eefsp}).  The  lowest order terms  in a derivative expansion
are \begin{equation}   1\;\;+ \;\;
\overleftarrow{ \frac\partial{\partial\zeta    }}\;   \theta(|\zeta|^2) \;
\overrightarrow{\frac\partial{\partial\bar\zeta}}\;\;+\;\;
\frac14 \biggl[ \overleftarrow{\frac{\partial^2} { \partial\zeta^2 }}
         \;\;  \overrightarrow{ \frac\partial { \partial\bar\zeta } }\;
\theta(|\zeta |^2)^2
 \overrightarrow{ \frac\partial { \partial\bar\zeta } }\; +\;
  \overleftarrow{\frac{\partial}{ \partial\zeta }}  \;  \theta(|\zeta |^2)^2\;
  \overleftarrow{\frac{\partial}{ \partial\zeta }}
   \;\;
  \overrightarrow{ \frac{\partial^2} { \partial\bar\zeta^2 } }
\biggr]  \;\;+\;\;\cdot\cdot\cdot
\label{afsvt}\end{equation}  This can also be regarded as   an expansion in
$\theta(|\zeta |^2)$, which is the covariant symbol of $\Theta$,
$\;\theta(|\zeta |^2) = <\zeta|\Theta (z z^\dagger)|\zeta>$. Since
$\theta(|\zeta |^2)\rightarrow 0$ in the limit of vanishing
 noncommutativity parameter $\Theta_0$,  the commutative limit
  is also the limit of small derivatives.   So for
  $\Theta_0\rightarrow 0$  we again get (\ref{pwp}), but now with the
  Poisson bracket
\begin{equation}
  \{ \phi,\psi\}(\zeta,\bar \zeta) =
-i \theta_{cl}(|\zeta|^2) \;
 \;\biggl( \frac{\partial\phi}{ \partial\zeta }\;\;
 \frac{\partial \psi}{ \partial\bar\zeta} \;-\;
 \frac{\partial\phi}{ \partial\bar\zeta }\;
\frac {\partial\psi}{ \partial\zeta} \biggr)\;,\end{equation}
 where $\;
 \theta_{cl}(|\zeta|^2)\; \equiv\; \lim_{\;\Theta_0\rightarrow 0}\; {
   \theta(|\zeta|^2)}\;/\;{\Theta_0}\;$.  From (\ref{lnl}) the limit
corresponds to large $n$ where $\Theta$ is slowly varying.
Then\footnote{For $r_0$ nonvanishing in this limit, we also need that
  $n_0\rightarrow\infty$, keeping $\Theta_0 n_0$ constant.}
\begin{equation}  \theta_{cl}(|\zeta|^2) =\frac 1{1+ \frac {r_0}{|\zeta|}}\;,
\end{equation}
which is defined away from the origin.  The result shows that the
commutative limit is a punctured plane.  At large distances   $|\zeta|
>>r_0$ from the
puncture one approaches the  Poisson structure associated with
the noncommutative plane. 

Now consider the operator $P_{n_0}= |n_0><n_0| $ which projects out
the ground state.  Its associated covariant symbol is $
|<\zeta|n_0>|^2$, which was identified in ref. \cite{nnsps}  as a
normalization constant in the expansion of the coherent states in
terms of the $|n>$ eigenstates.
Its asymptotic form was computed for small $\Theta_0$
in ref. \cite{nnsps} and found to be  a Gaussian with support
in a tiny region around the puncture:
\begin{equation}  |<\zeta|n_0>|^2 \;\rightarrow \; \exp\;\biggl\{ -\frac{|\zeta|^2
    + 2\sqrt{2}\; r_0 |\zeta|}{\Theta_0}\biggr\} \;,\qquad {\rm
    as}\qquad \Theta_0\rightarrow 0 \label{clsp}
\end{equation}
The  covariant symbol  of $P_{n_0}$  is thus defined at a point in the commutative limit.   In 
section 5 we shall see that all gauge invariant observables of
Chern-Simons theory on this noncommutative manifold  have this feature.

\section{Chern-Simons theory  on the  noncommutative plane}

We first consider Chern-Simons theory written on  ${\cal M}^{(0)}\times
{\mathbb{R}} $, where ${\mathbb{R}} $ corresponds to time, and show
that this is an empty theory. 
The  degrees of freedom for  noncommutative $U(1)$  Chern-Simons theory can be
taken to be a conjugate pair of potentials  $A$ and $ A^\dagger$. 
Under gauge transformations:
\begin{equation}
  A\rightarrow  i U^{\dagger}\nabla_z U+ U^{\dagger} AU \qquad
  A^\dagger\rightarrow  iU^{\dagger} \nabla_{z^\dagger}  U + U^{\dagger} 
A^\dagger U\;,
\end{equation}
 where $U$ is unitary
 function. 
  It is convenient to introduce 
$X=p+A$ and $ X^\dagger = p^\dagger + A^\dagger$ for they   transform covariantly:
$ X\rightarrow U^{\dagger} XU\; $, $\; X^\dagger\rightarrow U^{\dagger}
X^\dagger U\;.$ 
  The  field strength is
\begin{equation}
 { F}= i\nabla_{z} A^\dagger - i\nabla_{z^\dagger} A +[A, A^\dagger] = [X, X^\dagger] -[p, p^\dagger]\;, \label{nccurv}
\end{equation}
which then also transforms covariantly.  
  The  Chern-Simons Lagrangian can be written
\begin{equation}  L_{cs}=
 k \;{\rm Tr}\;\biggl[ \frac i2\Theta_0\;\biggl(D_{t}
X X^\dagger-X(D_{t} X) ^\dagger\biggl)\;
 +\; A_0 \biggr] \;,\label{csl}
\end{equation}
where  
$ D_t X=\dot X -i[A_0,X]\;, $ the dot denotes a time derivative and
Tr is the trace  over basis states in $ H^{(0)}$. 
  $A_0$ plays the role of a Lagrange
multiplier.  It is assumed to be hermitian and gauge transform as $ A_0  \rightarrow i U^\dagger
\dot U + U^\dagger A_0 U\;,
$ and so $ D_t X$ and its hermitian conjugate $(D_{t} X)^\dagger$
transform covariantly.  The constant $k$ is called the level.
Gauge invariance of  $\exp{ i\int_{ \mathbb{R} }
  dt\;L_{cs}} $ was   shown in refs. \cite{nair},\cite{Bak}  to lead to level
quantization  $ k= {\rm
  integer}\times \hbar$, and the integer
was identified in refs. \cite{sus},\cite{poly} with the
inverse of the filling fraction $\nu$ in the FQHE.

As with Chern-Simons theory on commutative ${\mathbb{R}}^3$, the above
 theory is empty.  This is easily seen in the canonical formalism. 
 The time derivative terms  in (\ref{csl}) define the Poisson
 structure.   The phase space is spanned by matrix elements   $
\chi_n^{\;m} =$   $ <n|X|m > $ and $  \bar\chi_n^{\;m} = <n| X^\dagger|m>  $,
 with Poisson brackets
\begin{equation} \{ 
\chi_n^{\;m} , \bar \chi_r^{\;s} \} =-\frac{i}{k\Theta_0}  \delta^m_r
 \delta^s_n \end{equation}
 The remaining terms  in the trace in (\ref{csl}) give the Gauss law constraints
\begin{equation}  G_n^{\;m} = <n|[X, X^\dagger]|m > +\Theta_0^{-1}\delta_n^m  = \chi_n^{\;r} \bar \chi_r^{\;m} - \bar  \chi_n^{\;r}  \chi_r^{\;m}
 +\Theta_0^{-1} \delta_n^m \approx 0 \;. \end{equation}  They are first class, and  from
\begin{eqnarray} ik\Theta_0 \;\{ 
\chi_n^{\;m}, G_r^{\;s}\} &=& \chi_r^{\;m}\delta_n^s -
\chi_n^{\;s}\delta_r^m \cr &&\cr  ik\Theta_0 \; \{ \bar
\chi_n^{\;m}, G_r^{\;s}\} &=&\bar \chi_r^{\;m}\delta_n^s - \bar
\chi_n^{\;s}\delta_r^m \end{eqnarray}  generate gauge transformations.  Since
every first class constraint eliminates two   phase space variables, no degrees of freedom remain after projecting to the
reduced phase space.

\section{Chern-Simons theory  on the  noncommutative plane with defect}

We now consider Chern-Simons theory  on  ${\cal M}^{(n_0)}\times
{\mathbb{R}} $.  The field strength $F$ involves  first order
 derivatives
$\nabla_{z} A^\dagger $ and $\nabla_{z^\dagger} A$.  For them to be
well defined   on  ${\cal M}^{(n_0)}$  we should impose the boundary
 conditions:
$ <n_0|A|n>=<n| A^\dagger|n_0> =0\;,\;\; \forall\; n\ge
  n_0\;.
$
Since $p$ and $ p^\dagger$
 are proportional to raising and lowering operators, respectively, we can also write
\begin{equation} <n_0|X|n>=<n| X^\dagger|n_0> =0\;,\qquad \forall\; n\ge n_0\label{bcox}\end{equation}
In order that these boundary conditions are preserved under gauge
 transformations we need the unitary matrices to satisfy
\begin{equation} <n_0|U^\dagger |a>=<a|U|n_0>=0 \;,\qquad  \forall\;a,b,...\ge n_0+1\label{rou} \end{equation}
Since gauge transformations are thereby restricted, not all phase
 space degrees of freedom in Chern-Simons theory can be gauged away,
 as was the case previously.

For the Chern-Simons Lagrangian we once again assume  (\ref{csl}),
 only now the trace is over a basis in
   $ H^{(n_0)}$.
Returning to the Hamiltonian formulation, and now imposing the
 boundary conditions (\ref{bcox}), one is left with  the following  phase space variables:
\begin{eqnarray}
\chi_a^{\;b} =\Theta_0 <a|X|b >   &\qquad &  \bar\chi_a^{\;b} =\Theta_0 <a| X^\dagger|b> \;,\cr&&\cr  
\psi_a = <a| X|n_0 > & \qquad & \bar  \psi^a = <n_0| X^\dagger|a>\;, \end{eqnarray}
where again $a,b,...>n_0$, and we  have re-scaled $\chi$ and $\bar\chi$
in order to later obtain the desired commutative limit. The nonzero Poisson brackets are
\begin{equation} \{ 
\chi_a^{\;b} , \bar \chi_c^{\;d} \} =-\frac{i}{k}{\Theta_0}\; \delta^b_c \delta^d_a \qquad
 \{ 
\psi_a , \bar \psi^b \} = -\frac{i}{k\Theta_0} \delta^b_a
 \label{pbcps}
 \end{equation}
For later convenience we also re-scale the Gauss law
constraints:
\begin{equation} G_a^{\;b} =\Theta_0^2 <a|[X, X^\dagger]|b > +\Theta_0\delta_a^b  = \chi_a^{\;c} \bar \chi_c^{\;b} - \bar  \chi_a^{\;c}  \chi_c^{\;b}+\Theta_0\delta_a^b
+\Theta_0^2\psi_a  \bar \psi^b  \approx 0
\label{tglc}\end{equation}
They generate gauge transformations which are consistent with (\ref{rou}):
\begin{eqnarray} ik\Theta_0^{-1} \; \{ 
\chi_a^{\;b}, G_c^{\;d}\} &=& \chi_c^{\;b}\delta_a^d -
  \chi_a^{\;d}\delta_c^b \cr  &&\cr
  ik\Theta_0^{-1} \; \{ \bar
\chi_a^{\;b}, G_c^{\;d}\}& =&\bar \chi_c^{\;b}\delta_a^d - \bar
\chi_a^{\;d}\delta_c^b \cr  &&\cr  ik\Theta_0^{-1} \; \{ 
 \psi_a, G_b^{\;c}\} &=&     \psi_b\delta_a^c \cr  &&\cr
   ik\Theta_0^{-1} \; \{ \bar
 \psi^a, G_b^{\;c}\}&= &-  \bar   \psi^c\delta_b^a \end{eqnarray}
From a counting argument alone the variables $
\chi_a^{\;b}$ and $  \bar\chi_a^{\;b}$ can be gauged away, leaving only
$ 
\psi_a $ and $  \bar  \psi^a$.  But the latter are not gauge
invariant.  Instead they transform as a vector and conjugate vector,
while  $
\chi_a^{\;b}$ and $  \bar\chi_a^{\;b}$ transform as tensors.  We can  construct
gauge invariant  observables of the form\footnote{Another set of gauge
  invariants are Tr $ (\bar \chi)^\alpha
(\chi)^\beta$ but their Poisson bracket algebra is ill-defined due to
  problems in defining
  the trace.}
\begin{equation} M_{(\alpha,\beta)} =- k \;\bar \psi (\bar \chi)^\alpha
(\chi)^\beta\psi  \label{defofM} \end{equation}
Their Poisson bracket algebra, and corresponding quantum algebra, were
computed in ref.  \cite{Pinzul:2002fi}
and found to be  a nonlinear
deformation of the classical $w_\infty$ algebra.
The classical $w_\infty$ algebra which generates area preserving diffeomorphisms is recovered in the commutative limit
 $\Theta_0\rightarrow 0 $,
\begin{equation}
 \{ M_{(\alpha,\beta)}, M_{(\rho,\sigma)} \}\rightarrow 
-i (\beta\rho -\alpha
 \sigma)\;
M_{(\alpha+\rho - 1, \beta + \sigma -1)}  \end{equation}

Finally we consider mapping the observables to the complex plane, and taking the commutative limit.   For this we once again utilize
coherent states of section 3, and construct covariant symbols of the relevant
operators.
For example, the covariant symbols associated with matrix elements $\psi_a = <a|
X|n_0 > $ and $ \bar  \psi^a = <n_0| X^\dagger|a>$ are
\begin{equation} \psi(\zeta,\bar\zeta) =<\zeta |a>\psi_a <n_0|\zeta> \;,\qquad 
\bar \psi(\zeta,\bar\zeta) =<\zeta |n_0>\bar \psi^a <a|\zeta> \;,
\end{equation} respectively.   Since the gauge invariant observables 
$ M_{(\alpha,\beta)}$ involve products of  $\psi_a $ and $ \bar
\psi^a$,  the covariant symbol associated with $ M_{(\alpha,\beta)}$  will involve star products of 
$\psi(\zeta,\bar\zeta)$ and $\bar \psi(\zeta,\bar\zeta)$.  They are
therefore determined by  $|<\zeta |n_0>|^2$ and its derivatives.   From
(\ref{clsp}) we then conclude that  in the commutative limit, all of the $w_\infty$ observables are
concentrated near the point singularity.  The area preserving
diffeomorphism generators
are then edge states.

\section*{Acknowledgments}

This work was supported in part  by the joint NSF-CONACyT grant E120.0462/2000.


\newpage

\begin{thebibliography}{0}




\bibitem{reviews}  See  for example, M.~R.~Douglas and N.~A.~Nekrasov,
Rev.\ Mod.\ Phys.\  {\bf 73}, 977 (2001).
           
\bibitem{wit} E. Witten, Commun. Math. Phys. {\bf 121}, 351 (1989);   M. Bos and V.P. Nair, Phys. Lett. {\bf B223}, 61 (1989);
  Int. J. Mod. Phys. {\bf A5}, 959 (1990); G.V. Dunne, R. Jackiw and
  C.A. Trugenberger, Ann. Phys. {\bf 194}, 197 (1989);
G. Moore and N. Seiberg, Phys.Lett. {\bf B220}, 422   (1989);  S. Elitzur, G. Moore, A. Schwimmer
 and N. Seiberg, Nucl.Phys. {\bf B326}, 108 (1989).

\bibitem{bbgs} A.P. Balachandran, G. Bimonte, K.S. Gupta, A. Stern,
 Int.J.Mod.Phys. {\bf A7} 4655 (1992); 5855 (1992).
\bibitem{sus}
 L. Susskind, hep-th/0101029. 

\bibitem{wit2}  E. Witten, Nucl. Phys. {\bf B311} 46 (1988).


\bibitem{nnsps}
A.~Pinzul and A.~Stern,
JHEP {\bf 0203}, 039 (2002)
[hep-th/0112220].



\bibitem{Pinzul:2002fi}
A.~Pinzul and A.~Stern,
Mod.\ Phys.\ Lett.\ A {\bf 18}, 1215 (2003)
[arXiv:hep-th/0206095].

\bibitem{IKS}
S.~Iso, D.~Karabali and B.~Sakita,
Phys.\ Lett.\ B {\bf 296}, 143 (1992);
A.~Cappelli, C.~A.~Trugenberger and G.~R.~Zemba,
Nucl.\ Phys.\ B {\bf 396}, 465 (1993).



\bibitem{poly}  A. P. Polychronakos, 
    JHEP {\bf 0011} 008 (2000);  {\bf 0104} 011 (2001);  hep-th/0106011.
\bibitem{us2}
A. Pinzul and A. Stern,  JHEP {\bf 0111} 023 (2001),  hep-th/0107179 ; A.~R.~Lugo, hep-th/0111064. 

\bibitem{Balachandran:2003vm}
A.~P.~Balachandran, K.~S.~Gupta and S.~Kurkcuoglu,
arXiv:hep-th/0306255.
\bibitem{Lizzi:2003ru}
F.~Lizzi, P.~Vitale and A.~Zampini,
arXiv:hep-th/0306247.


 \bibitem{gps}  G. Alexanian,  A. Pinzul and A. Stern, Nucl. Phys. {\bf B600},
531 (2001), hep-th/0010187.  




\bibitem{Voros}  F. Bayen, in {\it Group Theoretical Methods in Physics}, 
ed. E. Beiglb\"ock ,
et. al. [Lect. Notes Phys. {\bf 94}, 260 (1979)]; A. Voros, Phys. Rev. {\bf A 40},
6814 (1989).

\bibitem{Moyal} H. Groenewold, Physica (Amsterdam) {\bf 12}, 405
  (1946); J. Moyal, Proc. Camb. Phil. Soc. {\bf 45}, 99 (1949).


\bibitem{mmsz}  V.I. Man'ko,  G. Marmo, E.C.G. Sudarshan, F. Zaccaria, 
Physica Scripta {\bf 55}, 528 (1997).

\bibitem{nair} V.~P.~Nair and A.~P.~Polychronakos,
Phys.\ Rev.\ Lett.\  {\bf 87}, 030403 (2001).


\bibitem{Bak}
D.~Bak, K.~M.~Lee and J.~H.~Park,
Phys.\ Rev.\ Lett.\  {\bf 87}, 030402 (2001).


\end{thebibliography}
\end{document}